\documentclass[a4paper,showpacs,notitlepage,nobibnotes,pra,endfloats,preprint,footinbib,final,tightenlines,twocolumn,10pt]{revtex4}
\usepackage{amssymb}

\usepackage{amsmath}

\newtheorem{theorem}{Theorem}
\newtheorem{acknowledgement}[theorem]{Acknowledgement}

\input{tcilatex}

\begin{document}

\title{Decoherence and relaxation of qubits coupled to low- and
medium-frequency Ohmic baths directly and via a harmonic oscillator}
\author{Xian-Ting Liang}
\email{xtliang@ustc.edu;Tel:+86-574-87600783; Fax:+86-574-87600744}
\affiliation{Department of Physics and Institute of Modern Physics, Ningbo University,
Ningbo, 315211, China}

\begin{abstract}
Using the numerical path integral method we investigate the decoherence and
relaxation of qubits in spin-boson (SB) and spin-intermediate harmonic
oscillator (IHO)-bath (SIB) models. The cases that the environment baths
with low and medium frequencies are investigated. It is shown that the
qubits in SB and SIB models have the same decoherence and relaxation as the
baths with low frequencies. However, the qubits in the two models have
different decoherence and relaxation as the baths with medium frequencies.
The decoherence and relaxation of the qubit in SIB model can be modulated
through changing the coupling coefficients of the qubit-IHO and IHO-bath and
the oscillation frequency of the IHO.

\end{abstract}

\maketitle

\section{Introduction}

A two-level system, namely a qubit in today's modern physical language is a
most simple quantum physical model. If we only consider a closed qubit its
dynamics is simple enough and has been well known. However, in fact, there
is not any system being separated from its environment, in nature. The
interaction between the qubit and its environment should be considered when
we investigate the qubit's dynamics. The interaction make the open qubit
model become very complex and it may not be solved analytically exactly. In
order to investigate the model, in most cases, the environment is modeled
with a thermal bath which is considered being constructed with a set of
harmonic oscillators. A typical open qubit model is the well known
spin-boson (SB) model \cite{Weiss,RevModPhys_59_1,AnnPhys_149_374}. There is
another open qubit model in which the qubit couples to the coordinate $X$ of
a harmonic oscillator which we shall sometimes call the ``intermediate
harmonic oscillator'' (IHO) and which in turn is coupled to a bath \cite%
{JChemPhys_83_4491,AnnPhys_293_15,PhysRevLett_93_267005}. We call the model
the spin-IHO-bath (SIB) model.

It has be found that the two models have many physical correspondences. In
recent years the investigations of the two models have attached great
interest of scientists in felids of quantum information, electron transfer,
etc. \cite%
{ChemPhys_310_33,JChemPhys_118_179,JPhysChemA_103_9417,JChemPhys_102_4600}, %
\cite{ChemPhys_296_333,JModOpt_47_2905,EurPhysJB_45_405,PhysRevA_65_012309}.
These are motived by two exciting challenges, quantum computation and solar
energy utilization. It is believed that quantum computers may perform some
useful tasks more efficiently than their classical counterparts. Despite the
great promises of performing quantum computations, however, there are still
many practical difficulties to be resolved before quantum computers might
become available in future. One of the difficulties is that the qubit has
too short decoherence time, which is in fact a central impediment for
practical solid-state qubits to be taken as the cell of quantum computers.
Many significant results in the field, not only theoretical but also
experimental, have been achieved. Most of the theoretical research is based
on the SB and SIB models. On the other hand, the SB and SIB models can be
used to describe the electron transfer in chemical and biological molecules.
It has been recent found that the coherence is very important to electrons
for transferring energy in biological systems \cite%
{Nature_446_782,Science_316_1462}. So it is important to investigate the
decoherence and the relaxation of the electron spin (qubit).

In order to investigate the decoherence and relaxation of qubit in SB and
SIB models, essentially, one must understand the dynamics of the qubit \cite%
{AdvPhys_54_525,ChemPhys_296_135,JPhysSocJpn_75_082001}. If the qubit energy
splitting (denoted by $\Delta $ hereinafter) is not equal to zero, the two
models are not exactly solvable. However, they can be analyzed using
adiabatic renormalization in which a systematic weak damping approximation
must be used. They can also be investigated with some approximation methods
based on the perturbative scheme which also asks for the system (qubit)
weakly coupling to its environment. Many other methods \cite%
{PhysRevA_70_062106,ChemPhys_296_315,RepProgPhys_63_669,PhysRevE_72_056106,PhysRevE_56_334,PhysRevE_58_4288}
for solving the models have been proposed in recent years, most of which are
based on the Born-Markov approximation. However, it has been pointed out
that the use of the approximation is inappropriate at the large tunneling
amplitude and low temperatures. So it is important to find out some methods
to accurately estimate the dynamics of the qubits in the two models. Based
on the insight into the dynamics we may understand the decoherence and
relaxation better and may bring forward some schemes on how to suppress
them. An excellent, accurate numerical method based on the qusiadiabatic
propagator path integral (QUAPI) \cite%
{ChemPhysLett_221_482,JMathPhys_36_2430} may be a suitable tool for solving
the two models. To our problems we choose the iterative tensor
multiplication (ITM) algorithm for the numerical scheme. As Makri \cite%
{ChemPhysLett_221_482,JMathPhys_36_2430} addressed that the method is
non-Markovian and it can make the calculations accurate enough even at very
low temperatures, large tunneling amplitude and strong couplings for which
the Markovian approximation is unsuitable. In paper I \cite%
{ChemPhysLett_449_296} using the ITM algorithm we investigated the dynamics
and then the decoherence and relaxation of the qubit in the SB and SIB
models as the bath modes with higher frequencies than the oscillating
frequency of the IHO. In this paper by using the same ITM algorithm we shall
investigate the decoherence and relaxation of the qubit in low- and
medium-frequency baths. Where we consider one kind of bath, the Ohmic bath
interacting with the qubit. The low-frequency bath denotes that the Ohmic
bath modes with lower frequencies than the oscillating frequency of the IHO
and the medium-frequency bath refer that the Ohmic bath modes with the
frequencies approaching to the IHO's one.

\section{Models and dynamics}

The Hamiltonian of the SB model is%
\begin{equation}
H_{SB}=\frac{\hbar }{2}\left( \epsilon \sigma _{z}+\Delta \sigma _{x}\right)
+\tsum_{i}\left[ \frac{p_{i}^{2}}{2m_{i}}+\frac{1}{2}m_{i}\omega
_{i}^{2}\left( x_{i}+\frac{c_{i}\sigma _{z}}{m_{i}\omega _{i}^{2}}\right)
^{2}\right] .  \label{e1}
\end{equation}%
Suppose the bath has an Ohmic spectral density%
\begin{equation}
J_{ohm}\left( \omega \right) =\frac{\pi }{2}\tsum_{i}\frac{c_{i}^{2}}{%
m_{i}\omega _{i}}\delta \left( \omega -\omega _{i}\right) =\frac{\pi }{2}%
\hbar \xi \omega e^{-\omega /\omega _{c}}.  \label{e2}
\end{equation}%
Here, $\xi $ is the dimensionless Kondo parameter\ \cite%
{JChemPhys_102_4600,JChemPhys_102_4611} (the relationship of $\xi $ with the
friction coefficient $\eta $ is $\xi =2\eta /\pi \hbar $ \cite%
{PhysRevA_70_042101}), $\sigma _{i}$ $\left( i=x,\text{ }z\right) $ are the
Pauli matrix, and $\omega _{c}$ is the high-frequency cut-off of the bath
modes. This is a well-known quantum dissipation model and it has been widely
investigated \cite{Weiss,RevModPhys_59_1}.

If we consider the qubit coupling to the coordinate $X$ of a single IHO
which in turn is coupled to a bath, and if we let the couplings be linear,
the Hamiltonian of the SIB system reads%
\begin{eqnarray}
H_{SIB} &=&\frac{\hbar }{2}\left( \epsilon \sigma _{z}+\Delta \sigma
_{x}\right) +\frac{P^{2}}{2M}+\frac{1}{2}M\Omega _{0}^{2}\left( X+\lambda
\sigma _{z}\right) ^{2}  \notag \\
&&+\tsum_{i}\left[ \frac{p_{i}^{2}}{2m_{i}}+\frac{1}{2}m_{i}\omega
_{i}^{2}\left( x_{i}+\frac{\kappa c_{i}X}{m_{i}\omega _{i}^{2}}\right) ^{2}%
\right] ,  \label{e3}
\end{eqnarray}%
where $M$ and $P$ are the mass and momentum of the IHO, and the displacement 
$\lambda $ characterizes the coupling of the qubit to the IHO, and $\kappa
c_{i}$ are the coupling coefficients of the IHO to the bath modes. It is
shown that the system has a one to one map to the following system \cite%
{JChemPhys_83_4491}%
\begin{equation}
H_{SIB}=\frac{\hbar }{2}\left( \epsilon \sigma _{z}+\Delta \sigma
_{x}\right) +\tsum_{i}\left[ \frac{\tilde{p}_{i}^{2}}{2\tilde{m}_{i}}+\frac{1%
}{2}\tilde{m}_{i}\tilde{\omega}_{i}^{2}\left( \tilde{x}_{i}+\frac{\tilde{c}%
_{i}\sigma _{z}}{\tilde{m}_{i}\tilde{\omega}_{i}^{2}}\right) ^{2}\right] .
\label{e4}
\end{equation}%
According to paper I, as the high-frequency cut-off of the bath mode is not
infinite the effective spectral density is%
\begin{eqnarray}
J_{eff}\left( \omega \right)  &=&\frac{\pi }{2}\tsum_{i}\frac{\tilde{c}%
_{i}^{2}}{\tilde{m}_{i}\tilde{\omega}_{i}}\delta \left( \omega -\tilde{\omega%
}_{i}\right)   \notag \\
&=&\frac{\pi }{2}\lambda ^{2}\kappa ^{2}\xi \hbar \omega \frac{\Omega
_{0}^{4}}{\left( \omega ^{2}-\Omega _{0}^{2}\right) ^{2}e^{\omega /\omega
^{c}}+4\Gamma ^{2}\omega ^{2}e^{-\omega /\omega ^{c}}},  \notag \\
&&  \label{e5}
\end{eqnarray}%
where $\Gamma =\kappa ^{2}\eta /2M.$ The spectral density functions of $%
J_{ohm}\left( \omega \right) $ and $J_{eff}\left( \omega \right) $ versus
bath modes' frequencies are plotted in Fig. 1.

\begin{quotation}
\begin{eqnarray*}
&& \\
&&Fig.1
\end{eqnarray*}%
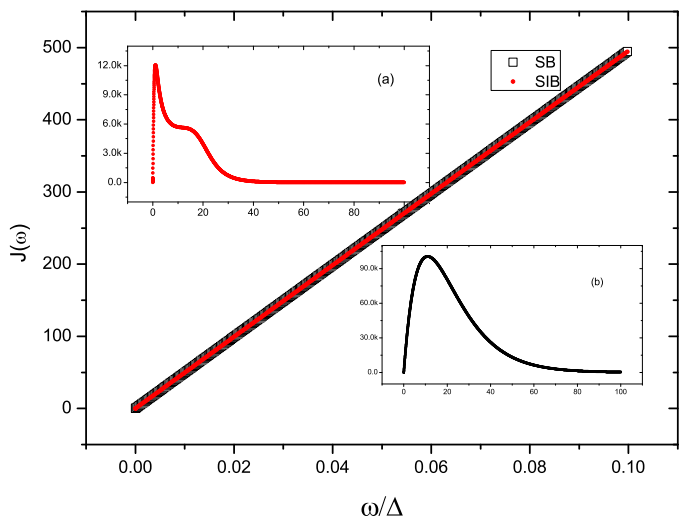
{\small Fig. 1: The spectral density functions }$J_{ohm}\left( \omega
\right) ${\small \ and }$J_{eff}\left( \omega \right) ${\small \ versus the
frequency }$\omega ${\small \ of the bath modes, where }$\Delta =5\times
10^{9}${\small \ Hz, }$\lambda \kappa =1,${\small \ }$\xi =0.01,${\small \ }$%
\Omega _{0}=10\Delta ${\small , }$T=0.01${\small \ K, }$\Gamma =2.6\times
10^{11}${\small \ Hz}$.$
\end{quotation}

Here, we set $\lambda \kappa =1$ and other parameters are same as in paper
I, namely, $\Delta =5\times 10^{9}$ Hz, $\xi =0.01,$ $\Omega _{0}=10\Delta $%
, $T=0.01$ K, $\Gamma =2.6\times 10^{11}$ Hz. From Fig.1 we see that the
spectral density functions have following characteristics. When the
frequencies of the bath modes are low ($0<\omega \leq 0.1\Delta $) the $%
J_{ohm}\left( \omega \right) $ and $J_{eff}\left( \omega \right) $ increase
linearly with the bath frequency $\omega $ and they are equivalent, see Fig.
1. When the frequencies of the bath modes are medium ($0.1\Delta <\omega
\leq 11\Delta $) the $J_{ohm}\left( \omega \right) $ and $J_{eff}\left(
\omega \right) $ behave in their different ways, see the insetted figures in
Fig. 1. In particular, in the model SIB, the IHO resonate with some mode of
the bath within the frequency limits. When the frequencies of the bath modes
are high ($11\Delta <\omega \leq 100\Delta $) the $J_{ohm}\left( \omega
\right) $ and $J_{eff}\left( \omega \right) $ decrease with the bath
frequency $\omega $. We call the bath modes with low- ($0<\omega \leq
0.1\Delta $), medium- ($0.1\Delta <\omega \leq 11\Delta $) and
high-frequency ($11\Delta <\omega \leq 100\Delta $) the low-, medium- and
high-frequency baths respectively. The range of frequencies for the low-,
medium- and high-frequency baths are plotted in Fig. 2 and they are
represented with Low F, Medium F and High F in the figure.

\begin{quotation}
\begin{eqnarray*}
&& \\
&&Fig.2
\end{eqnarray*}%
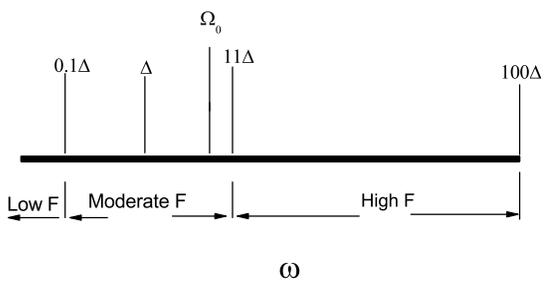
{\small Fig. 2: The sketch map on the low-, medium-, and high-frequency
baths.}
\end{quotation}

When the bath modes have high frequencies the dynamics of the qubit in SB
and SIB models has been investigated in paper I. In this paper we
investigate other two cases, namely, the cases of low- and medium-frequency
baths. The IHO is resonance with one of the modes of the medium-frequency
bath. The length of the memory times of the baths can be estimated by the
following bath response function%
\begin{equation}
\alpha \left( t\right) =\frac{1}{\pi }\int_{0}^{\infty }d\omega J\left(
\omega \right) \left[ \coth \left( \frac{\beta \hbar \omega }{2}\right) \cos
\omega t-i\sin \omega t\right] .  \label{e6}
\end{equation}%
Here, $\beta =1/k_{B}T$ where $k_{B}$ is the Boltzmann constant, and $T$ is
the temperature. It is shown that when the real and imaginary parts behave
as the delta function $\delta \left( t\right) $ and its derivative $\delta
^{\prime }\left( t\right) ,$ the dynamics of the reduced density matrix is
Markovian. However, if the real and imaginary parts are broader than the
delta function, the dynamics is non-Markovian. The broader the Re$[\alpha
\left( t\right) ]$ and Im$[\alpha \left( t\right) ]$ are, the longer the
memory time will be. The broader the Re$[\alpha \left( t\right) ]$ and Im$%
[\alpha \left( t\right) ]$ are, the more serious the practical dynamics will
be distorted by the Markovian approximation. The memory time of the
effective bath is affected by $\Gamma .$ The larger the $\Gamma $ is, the
shorter the memory time of the effective bath will be. Clearly, the value of
the $\Gamma $ may be different according to the difference of the physical
systems. For example, when the persistent-current qubit is measured by a dc
SQUID, the system can be modeled by Eq. (\ref{e4}) with Eq. (\ref{e5}), here 
$\Gamma =1/R_{s}C_{s}.$ Typically, $R_{s}=100$ $\Omega ,$ $C_{s}=5\ $pF, so $%
\Gamma \sim 10^{11}$ Hz, see Ref. \cite{PhysRevB_65_144516}. Similar to
paper I we set $\Gamma =2.6\times 10^{11}$ Hz in this paper$.$ In Fig. 3 we
plot the Re$[\alpha \left( t\right) ]$ and Im$[\alpha \left( t\right) ]$ of
the low-frequency bath as (a) $J\left( \omega \right) =J_{ohm}\left( \omega
\right) $, and (b) $J\left( \omega \right) =J_{eff}\left( \omega \right) $
and the medium-frequency bath as (c) $J\left( \omega \right) =J_{ohm}\left(
\omega \right) $, and (d) $J\left( \omega \right) =J_{eff}\left( \omega
\right) .$ It is shown that the memory times in above four cases are all
about $\tau ^{m}\approx 1.5/\Delta .$

\begin{quotation}
\begin{eqnarray*}
&& \\
&&Fig.3
\end{eqnarray*}%
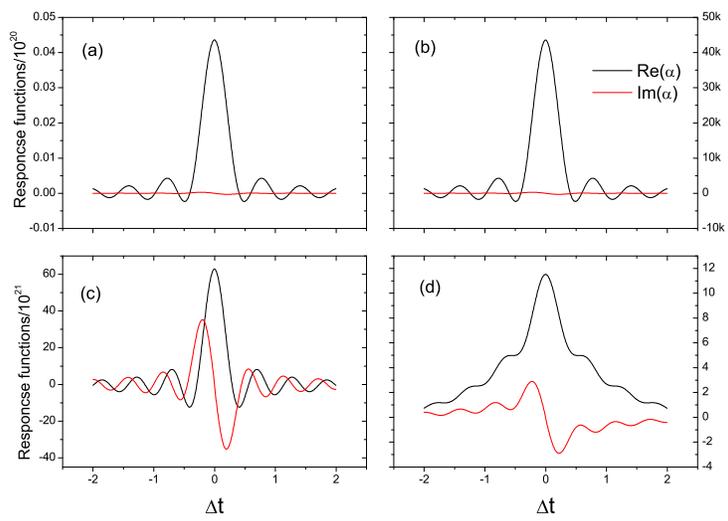
{\small Fig. 3: The response functions of the Ohmic bath in (a) low and (c)
medium frequencies and effective bath in (b) low and (d) medium frequencies.
The parameters are the same as in Fig. 1. The cut-off frequencies for the
two cases are taken according to Fig. 2.}
\end{quotation}

The dynamics of the qubit is characterized by the time evolution of the
reduced density matrix, obtained after tracing out the bath degrees of
freedom, i.e.,%
\begin{equation}
\rho \left( s^{\prime \prime },s^{\prime };t\right) =\text{Tr}%
_{bath}\left\langle s^{\prime \prime }\right| e^{-i\mathcal{H}t/\hbar
}R\left( 0\right) e^{i\mathcal{H}t/\hbar }\left| s^{\prime }\right\rangle .
\label{e7}
\end{equation}%
Thoughout this paper we assume that the interaction between system and bath
is turned on at $t=0$, such that the initial density matrix factorizes into
its system and bath components, and the bath is initially at thermal
equilibrium \cite{JChemPhys_102_4611,PhysRevA_70_042101}:%
\begin{equation}
R\left( 0\right) =\rho \left( 0\right) \otimes \rho _{bath}\left( 0\right) ,
\label{e8}
\end{equation}%
where $\rho \left( 0\right) $ and $\rho _{bath}\left( 0\right) $ are the
initial states of the qubit and bath. Here, we calculate the reduced density
matrix $\rho (t)$ by using the well established ITM algorithm derived from
the QUAPI. This algorithm is a numerically exact one and is successfully
tested and adopted in various problems of open quantum systems \cite%
{JChemPhys_102_4600,JChemPhys_102_4611,PhysRevE_62_5808}. For details of the
scheme, we refer to previous works \cite%
{ChemPhysLett_221_482,JMathPhys_36_2430}. The QUAPI asks for the system
Hamiltonian splitting into two parts $H_{0}$ and $H_{env}.$ Here, we take $%
H_{0}=\frac{\hbar }{2}\left( \epsilon \sigma _{z}+\Delta \sigma _{x}\right) $
and $H_{env}=H_{SB}-H_{0}$, or $H_{env}=H_{SIB}-H_{0}$. In order to make the
calculations converge we use the time step $\Delta t=0.5/\Delta ,$ which is
smaller than the characteristic times of the qubits in the systems.

\section{Decoherence and relaxation}

The decoherence is in general produced due to the interaction of the quantum
system with other systems with a large number of degrees of freedom, for
example the devices of the measurement or environment. To measure the
decoherence one may use the entropy, the first entropy, and many other
measures, such as the maximal deviation norm, etc. (see for example Refs. %
\cite{JStatPhys_110_957,PhysRevA_69_032311}). However, essentially, the
decoherence of an open quantum system is reflected through the decays of the
off-diagonal coherent terms of its reduced density matrix \cite%
{PhysRevB_72_245328}. The decoherence time denoted by $\tau _{2}$ measures
the time of the initial coherent terms to their $1/e$ times, namely, $\rho
_{i}\left( n,m\right) \overset{\tau _{2}}{\rightarrow }\rho _{f}\left(
n,m\right) =\rho _{i}\left( n,m\right) /e.$ Here, $n\neq m$, and $n,$ $m=0$
or $1$ for qubits. In the following, we investigate the decoherence of the
qubit in SB and SIB models via directly calculating the evolutions of the
off-diagonal coherent terms, instead of some measure of the decoherence.
Similar to the decoherence, the relaxation of the qubit can also be
investigated with calculating the evolutions of the diagonal elements of the
reduced density matrix. The relaxation time is denoted by $\tau _{1},$ which
measures the time of an initial state to the final thermal equilibrium state
through estimating the diagonal terms of the reduced density matrix, namely, 
$\rho _{i}\left( n,n\right) \overset{\tau _{1}}{\rightarrow }e^{-E_{n}\beta
}.$ In the following calculations we assume that the initial state of the
environment is $\rho _{bath}\left( 0\right) =\prod\nolimits_{k}e^{-\beta
M_{k}}/$Tr$_{k}\left( e^{-\beta M_{k}}\right) $ and initially the qubit in
its maximal coherent state. Here$,$ $M_{k}=\hbar \omega _{k}b_{k}^{\dagger
}b_{k}$ where $b_{k}^{\dagger }$ ($b_{k})$ are the create (annihilate)
operators of the $k-th$ mode for the environment. As calculating the
off-diagonal element $\rho _{12}$ we let $\epsilon =10\Delta $ which can
make the $\rho _{12}$ decay stably. If $\epsilon \rightarrow \Delta $ the $%
\rho _{12}$ will decay with some oscillations, which may affect our
judgement on decoherence times. The closer the two parameters are, the more
strongly the matrix elements will oscillate. When we calculate $\rho _{11}$
we choose parameters $\epsilon =\Delta $ because the oscillations of the $%
\rho _{11}$ do not affect our judgement on relaxation times from the
figures. If we choose parameters $\epsilon =10\Delta $ other than $\epsilon
=\Delta $ in calculating the $\rho _{11},$ the relaxation of the qubit
cannot be shown clearly in the figures in a short time because the time $%
\tau _{1}$ is greatly larger than the time $\tau _{2}$ in our problems. So,
the reader should note that, the increase of the $\Delta \ $and $\epsilon $
will shorten the decoherence and relaxation times, and the decoherence time $%
\tau _{2}$ and the relaxation time $\tau _{1}$ in following figures are not
comparable because they are plotted in different two sets of the parameters.
In the ITM scheme, one should at first choose the $k_{\max }$ and assure
that $k_{\max }\Delta t$ is larger than the effective memory time $\tau ^{m}$
of the baths. Fig. 3 and numerical tests tell us that the calculations are
in fact convergent as $k_{\max }\geqslant 3$. It is known that the qubit
will show different decoherence and relaxation when it has different initial
states. In the low- and medium-frequency baths, the decoherence and
relaxation of the qubits in their different initial states are different,
which is similar to the cases of high-frequency bath (see Fig. 3 of paper
I), we do not plot the evolutions of $\rho _{12}$ and $\rho _{11}$ in
different initial states in this paper. Fig. 4 plots the evolutions of the $%
\rho _{12}$ (below) and $\rho _{11}$ (up) of the qubit in low-frequency bath
in SB and SIB models. It is shown that the qubit in SB and SIB models has
almost same decoherence and relaxation in this case.

\begin{quotation}
\begin{eqnarray*}
&& \\
&&Fig.4
\end{eqnarray*}%
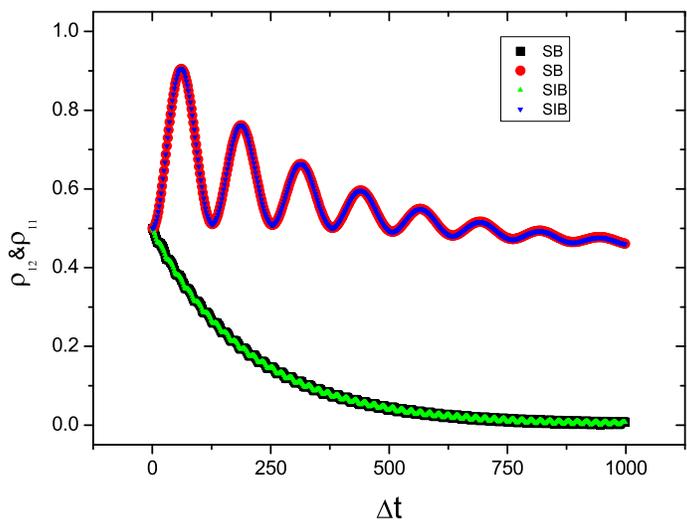
{\small Fig. 4: The evolutions of reduced density matrix elements }$\rho
_{12}${\small \ (below) and }$\rho _{11}${\small \ (up) in SB and SIB models
in low-frequency bath. The parameters are the same as in Fig. 1.}%
\begin{eqnarray*}
&& \\
&&Fig.5
\end{eqnarray*}%
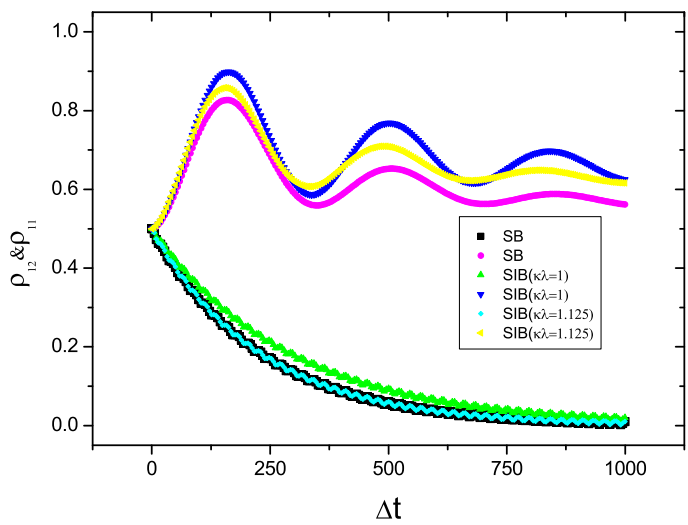
{\small Fig. 5: The evolutions of reduced density matrix elements of }$\rho
_{12}${\small \ (below) and }$\rho _{11}${\small \ (up) in SB and SIB models
in the medium-frequency bath (}$\lambda \kappa =1,${\small \ or }$1.125$%
{\small ). The other parameters are the same as in Fig. 1.}%
\begin{eqnarray*}
&& \\
&&Fig.6
\end{eqnarray*}%
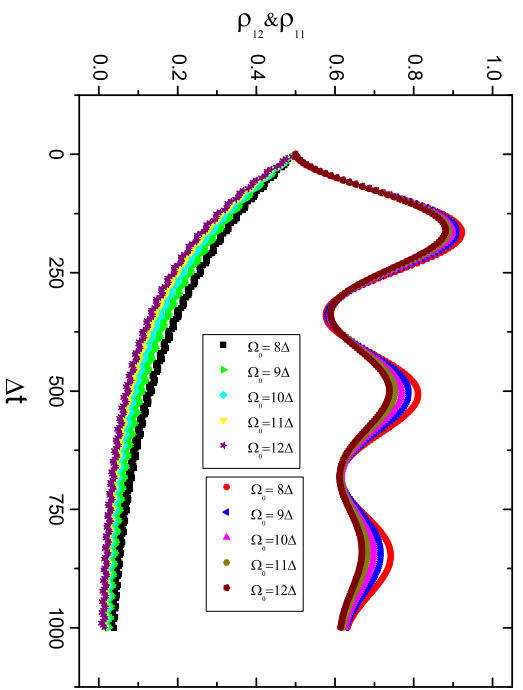
{\small Fig. 6: The evolutions of reduced density matrix elements of }$\rho
_{12}${\small \ (below) and }$\rho _{11}${\small \ (up) in SIB model in
medium-frequency bath in different values of }$\Omega _{0}${\small , the
other parameters are the same as in Fig. 1.}
\end{quotation}

Fig. 5 plots the evolutions of the $\rho _{12}$ (below) and $\rho _{11}$
(up) of the qubit in medium-frequency bath in SB and SIB models. It is shown
that as $\kappa \lambda =1,$ the qubit in SIB model has longer decoherence
and relaxation times than the qubit in SB model has. As $\kappa \lambda $
increase to $1.125$ the qubit in SIB model has almost same decoherence and
relaxation times to the qubit in SB model has, but the $\rho _{11}$ decays
to a bigger equilibrium value. Fig. 6 plots the $\rho _{12}$ (below) and $%
\rho _{11}$ (up) of the qubit in SIB model in different $\Omega _{0}$ of the
IHO in medium-frequency bath. It is shown that the decoherence and
relaxation times of the qubit in the SIB model increase with the decrease of
the oscillation frequency $\Omega _{0}$ of the IHO, which is similar to the
case that the bath has high frequencies.

\section{Conclusions and discussions}

In this paper we have investigated the decoherence and relaxation of a qubit
coupled to an Ohmic bath directly and via an IHO. In our investigations, we
fix the tunneling splitting $\Delta $ of the qubit. Two kinds of cases
complementing paper I are discussed. The first is that the frequencies of
the bath modes are low, and the IHO is far off the resonance to the bath
modes, i.e. the oscillation frequency of the IHO is larger than the
high-frequency cut-off $\omega _{c}$ of the bath modes. The second is that
the frequencies of the bath modes are medium and in the case the IHO
resonate with some mode of the bath. Here, we suppose that, in SIB model,
the value of the damping coefficient $\Gamma $ of the bath to the IHO is
intermediate as supposed in paper I, which make the bath has shorter memory
time and it is suitable for our using the ITM based on the QUAPI. By using
the ITM numerical scheme we calculated the evolutions of the reduced density
matrix elements of qubit in the two models. Some new results are obtained.
(1) When the frequencies of the bath modes are low ($0<\omega \leq 0.1\Delta 
$) the qubit in the SB and SIB models has almost the same decoherence and
relaxation times. Here, we set $\lambda \kappa =1$ other than $\lambda
\kappa =1050$ in paper I. If the damping of the bath to the IHO or (and) the
IHO to the qubit increase, namely $\lambda \kappa $ increase, the
decoherence and relaxation times of the qubit in SIB model will decrease,
and vice versa. (2) When the frequencies of the bath modes are medium ($%
0.1\Delta <\omega \leq 11\Delta $) and $\lambda \kappa =1,$ the qubit in the
SIB model will have longer decoherence and relaxation times than it has in
the SB model. As $\lambda \kappa $ increase to $1.125$, the qubit in SIB
model will have same decoherence and relaxation times to the qubit in SB
model has, but the $\rho _{11}$ decays to a bigger equilibrium value. (3)
The decoherence and relaxation times of the qubit in the SIB model increase
with the decrease of the oscillation frequency $\Omega _{0}$ of the IHO. (4)
As point out in paper I that the decoherence and relaxation times of the
qubit in the SB and SIB models will increase with the decrease of the $%
\epsilon $ and $\Delta $, which has also not been plotted in the paper. The
longer decoherence and relaxation times are necessary for not only the
qubits for making the quantum computers but also the electrons for
transferring energy in biological systems. In order to make the qubits or
electrons in the SIB model with longer decoherence and relaxation times we
may try to make the $\Omega _{0}$ and $\lambda \kappa $ smaller.

If the qubits in SB and SIB in full-frequency ($0<\omega \leq 100\Delta )$,
or in low- and medium-frequency ($0<\omega \leq 11\Delta $), or medium- and
high-frequency ($11\Delta <\omega \leq 100\Delta $) baths, how about the
dynamics of the qubits? This is an interesting problem. But, because the
effective bath in the frequency spectra has so long memory times that the
ITM algorithm in fact can not be applied for investigating the dynamics of
the qubit in these cases. So, some new methods is expected for the goal. In
paper I and this paper, we suppose that the qubit couple to the Ohmic bath,
or the IHO couple to the Ohmic bath. If we set the qubit or the IHO couple
to other baths, for example, sub-Ohmic and super-Ohmic baths, the method
used in the two papers is also valid, but in order to make the calculations
convergence one should recalculate the memory times of the baths and choose
the $k_{\max }$ again, so that $k_{\max }\Delta t\gtrsim \tau ^{m}$.

\begin{acknowledgement}
This project was sponsored by National Natural Science Foundation of China
(Grant No. 10675066) and K.C.Wong Magna Foundation in Ningbo University.
\end{acknowledgement}

\end{document}